\documentclass[journal]{IEEEtran}
\usepackage{cite}

\ifCLASSINFOpdf
  \usepackage[pdftex]{graphicx}
\else
  \usepackage[dvips]{graphicx}
\fi
\usepackage{amsmath}
\usepackage{algorithmic}

\usepackage{multirow}
\usepackage{booktabs}
\usepackage{subfigure}
\usepackage{color,xcolor}

\begin{document}
%
\title{Physiological-Physical Feature Fusion for Automatic Voice Spoofing Detection}
%
%
%

\author{Junxiao~Xue,
        Hao~Zhou,
        Yabo~Wang
\thanks{J. Xue was with the School of Software,  Zhengzhou University.}
\thanks{H. Zhou are with Zhengzhou University.}
\thanks{Y. Bo are with Westlake University.}
\thanks{Manuscript received April 19, 2005; revised August 26, 2015.}}

%
%

\markboth{Journal of \LaTeX\ Class Files,~Vol.~14, No.~8, August~2015}%
{Shell \MakeLowercase{\textit{et al.}}: Bare Demo of IEEEtran.cls for IEEE Journals}
%



\maketitle

\begin{abstract}
Speaker verification systems have been used in many production scenarios in recent years. Unfortunately, they are still highly prone to different kinds of spoofing attacks such as voice conversion and speech synthesis, etc. In this paper, we propose a new method base on physiological-physical feature fusion to deal with voice spoofing attacks. This method involves feature extraction, a densely connected convolutional neural network with squeeze and excitation block (SE-DenseNet), multi-scale residual neural network with squeeze and excitation block (SE-Res2Net) and feature fusion strategies. We first pre-trained a convolutional neural network using the speaker's voice and face in the video as surveillance signals. It can extract physiological features from speech. Then we use SE-DenseNet and SE-Res2Net to extract physical features. Such a densely connection pattern has high parameter efficiency and squeeze and excitation block can enhance the transmission of the feature. Finally, we integrate the two features into the SE-Densenet to identify the spoofing attacks. Experimental results on the ASVspoof 2019 data set show that our model is effective for voice spoofing detection. In the logical access scenario, our model improves the tandem decision cost function (t-DCF) and equal error rate (EER) scores by 4\% and 7\%, respectively, compared with other methods. In the physical access scenario, our model improved t-DCF and EER scores by 8\% and 10\%, respectively.
\end{abstract}


%
\IEEEpeerreviewmaketitle

\section{Introduction}
%
%
%
%
\IEEEPARstart{S}{mart} voice assistants have come into our lives. At present, many smart devices will integrate smart voice assistants, such as Samsung's Bixby, Apple's Siri, and Microsoft's Cortana. These smart voice assistants can provide personalized services to users by recognizing their voices. Automatic speaker verification (ASV) is the main technique used to recognize speaker voice \cite{reynolds2002overview}. It is a convenient biometric person authentication system that recognizes the speaker’s identification based on speech recordings. With the development of deep neural network technology, the automatic speaker verification system has achieved perfect effects and has been applied to many life and production scenarios, such as intelligent voice assistant, secure building access, e-commerce and speech emotion recognition, etc. However, ASV systems are still subject to many attacks \cite{unknown}. The most common attacks are voice conversion (VC), text to speech (TTS), and replay attack. VC can convert the speech without losing the target speaker’s distinct characters, which is one of the most accessible methods of attack. TTS can intelligently convert text into natural speech, and the sound is smooth, so that the listener feels natural when listening to information, without the coldness and numbness of computer-generated speech. Replay attack means that the attacker sends a received message to the target host to achieve the purpose of deceiving the system. It is mainly used in the process of identity authentication to destroy the correctness of identity authentication. In ASV system, replay attack is when the attacker records the voice of the target speaker and tries to pass the ASV system authentication as the target speaker. Due to the rapid development of deep learning technology, these attacks can already be very similar to real speech. This is a great challenge to the ASV system.

In order to overcome the above challenges, we need a system with excellent performance to distinguish real speech and spoofing speech. In the beginning stage, different evaluation standards are used to carry out the study on different datasets, and the results cannot be compared. In order to build a community with standard datasets and evaluation metrics, a series of anti-spoofing competitions were born, one of the most famous is the automatic speaker verification spoofing and countermeasures (ASVspoof) challenge. They raised this serious question back in 2013. In versions of ASVspoof 2015 and ASVspoof 2017, the spoofing countermeasures (CM) for synthetic speech and replay speech are designed respectively, but all types of spoofing attacks are centralized for the first time in ASVspoof 2019. It is divided into two scenarios, logical access (LA) and physical access (PA). LA focused on TTS and VC, while PA was designed to develop countermeasures capable of discriminating between genuine audios and replay ones. For ASV systems, it is difficult to defend against unknown attacks. To solve this problem, ASVspoof 2019 divided the data set into three parts: training set, development set and evaluation set. The evaluation set contains attacks that did not occur in the other two sets. So what succeeds in the development set can still do badly in the evaluation set. This requires our method to be universal to the unknown attack, and build a wide range of applicable countermeasures to make our system more robust.

At present, the research on anti-spoofing countermeasures of ASV mainly focuses on two aspects: one is to research the method of extracting features from speech; the other is to Looking for more efficient classifiers \cite{suthokumar2018modulation, alam2018boosting, sailor2018auditory}, The main classifier types are statistical modeling \cite{delgado2018asvspoof, adiban2017sut, khoury2014introducing} and deep neural networks (DNNs) \cite{chen2017resnet, lai2019attentive, valenti2018end, lai2019assert, monteiro2019end}. Previous researches have performed well in ASVspoof 2019. For example, the authors of \cite{monteiro2019end} proposed a light convolutional gated recurrent neural network based on gated recurrent units (GRU) by fusing Light RNN and CNN \cite{wu2018light}. They use this network as a deep feature extractor to extract deeper feature, so that assisting in the training of the classifier. The authors of \cite{chen2017resnet} built five DNN models and used different forms of feature to detect spoofing attacks. Features included a unified feature map, acoustic features, and whole utterances, which were fed into five DNN models based on variants of squeeze-excitation networks (SENets) \cite{hu2018squeeze} and ResNets \cite{he2016deep}.

In this paper, Using the advantages of multi-feature fusion and based on the success of deep neural networks in this task, we propose two different strategies for PA and LA. For the LA, we propose a new method, including feature extraction, a densely connected convolutional neural network with squeeze and excitation block , and feature fusion strategies, to deal with the spoofing attacks. Inspired by the success of Speech2Face model \cite{Oh2020Speech2Face}, which can translate speech into face feature. We use the fine-tuned Speech2Face model as one of the modules of our method. This module is to input a speech and get 4096-D face features. We use it as a depth feature extractor to extract face feature. We use SE-Densenet as another depth feature extractor to extract 128-D speech feature. Finally, we fused the two features and put them into the classifier. For the PA, we still use Speech2Face Model to capture face features. But we used a new neural network, SE-Res2Net, to extract speech features. In addition, the fusion strategy we used was also changed. In the PA, we adopted the back-end fusion strategy. Experiments show that this fusion strategy is more effective for PA. To address the challenges in ASVspoof 2019, we compared the performance of our models with the state-of-the-art models.

The major contributions of this work are as follows:

\noindent1. We design a novel convolutional network for audio spoofing detection. The network contains dense connection and squeeze and excitation block. As far as we know, this is the first work for the audio spoofing detection task using dense connection combined with squeeze and excitation block. Such a densely connected pattern has high parameter efficiency and squeeze and excitation block can enhance the transmission of the feature.

\noindent2. We combine physiological and physical features for audio spoofing detection. Through the network we designed, we can extract face feature and speech feature, and fuse two features. Our experiments show that this fusion feature can improve the performance of our model.

\noindent3. The developed network model achieves better results in the challenges of ASVspoof 2019 than state-of-the-art methods. For example, in the logical access scenario, our model improves the tandem decision cost function and equal error rate scores by 28\% and 11\%, respectively, compared with state-of-the-art methods.

The other parts of this paper are organized as follows. In section 2, we introduce the relevant technologies used in this paper. The model we proposed is introduced in section 3. Section 4 gives our specific experimental steps and results, and finally we draw a conclusion in Section 5.

\section{Relate Work}
\subsection{ Voice Spoofing Detection}

Synthetic speech is the main way of spoofing attack. ASVspoof proposes an anti-spoofing research based on ASV system. The main goal of this research is to protect ASV systems away from the threat of spoofing attack. The early work of fake audio detection focuses on extract features from audio. For example, \cite{2017ResNet} proposed a fusion model for automatic spoofing detection. They used different methods to extracting audio feature, and research the effect of these methods on automatic spoofing detection. They used the constant Q cepstrum coefficient (CQCC) and the Mel frequency cepstrum coefficient  (MFCC) to extract feature from the audio. Then they used GMM, ResNet, and DNN as classifier. They research the effects of different combinations. Similarly, \cite{DBLP:journals/corr/abs-1904-05576} proposed anti-spoofing Systems. They tried to use different feature extraction methods to influence the effect of the systems. They use different methods to extract feature, for example, constant Q transform  (CQT), fast fourier transform (FFT), discrete cosine transform  (DCT). Then they look for which works best for the systems.  Besides, \cite{2019The} proposed a robust anti-spoofing System for the ASVspoof 2019 challenge. They use log-CQT, log mel spectrogram (LMS), and Phase feature to extract feature, and use ResNet, ResNet with i-vector, LightCNN with multi-task outputs, Context Gate CNN as classifier. They explore the detection effects of different feature extraction methods and different model combinations. To sum up the above research, we can divide fake audio detection models into the following categories:

\subsubsection{Voice spoofing detection based on gaussian mixture model} The Gaussian model can break things down into one model of several models based on gaussian probability density functions and accurately quantify things. The original speech data will become feature sequences after short-time Fourier transform or cepstrum. Under the condition of ignoring the timing information, the gaussian mixture model is very suitable for fitting such speech features. Therefore, ASVspoof 2019 uses a common gaussian mixture model as a baseline system for spoofing audio detection.

\subsubsection{Voice spoofing detection based on convolutional neural network} In general, audio spoofing detection is usually based on matrix containing speech information, which has the characteristics of structure. In order to improve the accuracy of spoofing audio detection, it is necessary to overcome the diversity of speech signals. Convolutional neural networks can provide translational invariant convolutional in time and space. When applying the idea of convolutional neural network to acoustic modeling of audio spoofing detection, convolutional invariance can be used to overcome the diversity of speech signals themselves.From this perspective, we can think of the matrix derived from speech as an image matrix and process speech in the same way that images are processed, such as deep convolutional neural networks. This can effectively improve the accuracy of audio spoofing detection.

\subsubsection{Voice spoofing detection based on recurrent neural network} Recurrent neural network is a kind of neural network with memory function, which is suitable for modeling sequence data. It has achieved success in speech recognition, natural language processing, and other fields. Compared with the convolutional neural network, the recurrent neural network has the advantages of accepting the sequence data of variable length as input and has the memory function. Because the context correlation of speech fragments is very strong, the recurrent neural network is very helpful for the detection of spoofing audio.

These systems have achieved good results and achieved excellent results in spoofing detection of known systems, but their detection of spoofing attacks on unknown systems still falls short. We think it has to do with the input, the model we proposed in this paper not only take audio feature, but also extract physiological feature from audio. Then we fuse physiological feature and audio feature as input to the classifier.

\subsection{Speech2Face Model}


In natural face images, the large variability of facial expressions, head poses, occlusion and lighting conditions makes it very important to design and train a speech-face model. For example, the method of directly returning from input voice to image pixels is not feasible; such a model must learn to extract many irrelevant changes from the data, and extract some significative internal representations of the face. This in itself is a challenging task.


To overcome these challenges, \cite{Oh2020Speech2Face} A trained their model to return to the low-dimensional middle face representation. More specifically, they use the VGG-Face model, which is a face recognition model pre-trained on a large face data set \cite{parkhi2015deep}, and extract 4096-D face features from the second layer of the network. These face features contain enough information to reconstruct the corresponding facial image, and are robust to many of the above-mentioned changes.


Their speech2face model mainly consists of the following parts: 1) The voice encoder uses a complex speech spectrogram as input to predict the low-dimensional face features corresponding to the relevant face; 2) The face decoder, which uses facial features as input to generate face images in standard forms (frontal and neutral expressions). During the training process, the face decoder is fixed, and we only train the voice encoder to predicts face features. The voice encoder is a model they designed and trained, however the face decoder uses the model previously proposed by \cite{cole2017synthesizing}.


The voice encoder module is actually a convolutional neural network, which converts the spectrogram of the short input speech into pseudo-face features, and then inputs the pseudo-face features into the face decoder to reconstruct the face image. The blocks of convolutional layer, batch normalization \cite{ioffe2015batch} and ReLU alternate with the max-pooling layer. The max-pooling layer only pools along the time dimension of the spectrogram, while retaining frequency information. This is to preserve more sound features because they are better contained in frequency content, while language information usually lasts longer  \cite{hsu2017unsupervised}. At the end of these blocks, we apply average pooling along the time dimension. This allows us to efficiently aggregate information over time and makes the model suitable for input speech of different durations. Then, these pooled features are fed into two fully connected layers to generate a 4096-D face feature.


The goal of the face decoder is to reconstruct face images from low-dimensional face features. They choose to eliminate any irrelevant changes (posture, lighting, etc.) while preserving facial attributes. To this end, they used the face decoder model of Cole et al. \cite{cole2017synthesizing} to reconstruct a standard facial image of a frontal face containing only neutral expressions. They used the same face features extracted from the VGG-Face model as input to the face decoder to train this model. The model is trained separately and remains fixed during the training of the voice encoder.


\subsection{Feature Extract}

Feature extraction is to extract useful information from speech. A good feature extraction method can improve the efficiency of our system. At present, the common methods are MFCC, LPC, LPCC, LFCC, etc \cite{sahidullah2015comparison}. These methods have been widely used in previous research and achieved good results.  This makes them highly reliable and acceptable. \cite{davis1980comparison} propose the Mel frequency cepstrum coefficient (MFCC), which is a cepstrum parameter extracted in the Mel scale frequency domain. It has been widely used in automatic speech recognition. Spectrogram is a time-frequency graph based on the characteristics of speech signal, which can reflect the dynamic spectrum feature of audio data \cite{kingsbury1998robust}. Spectrogram is a kind of 2-dimension spectrum, which represents the graph that the spectrum of speech signal changes with time. The vertical axis is frequency and the horizontal axis is time. The shade of its tone indicates the energy intensity of the corresponding frequency at the corresponding time, so it can well display the characteristics of frequency and time-domain waveform. LPC is mainly used in speech processing, it is a tool that represents the spectrum limits of digital information signals in compressed form based on linear prediction model information. It is one of the most effective speech analysis techniques and one of the most useful methods to encode high quality speech at low speed. It also can provide very accurate speech parameter prediction, because it simulates the human vocal tract, so it has strong linearity and robustness. LPCC is the cepstrum coefficient obtained from the spectrum envelope calculated by LPC. It is the coefficient of the Fourier transform of the logarithmic amplitude spectrum of LPC. LPCC is a commonly used analysis method in the field of speech processing because it can perfectly represent the speech waveform and features with limited features. The LFCC is a form of cepstrum. It is the result of taking the logarithm of the spectrum of the input speech signal and then fourier transform it. It is called linear cepstrum because these coefficients are calculated by a linear filter bank with better resolution at higher frequencies. In this paper, spectrogram and LFCC is used as a method to extract audio feature, because they contains a lot of time and space information, which have great help to our classification task \cite{fu2013optimizing}.

\section{methodology}
\begin{figure*}[h]
	\centering
	\includegraphics[scale=0.17]{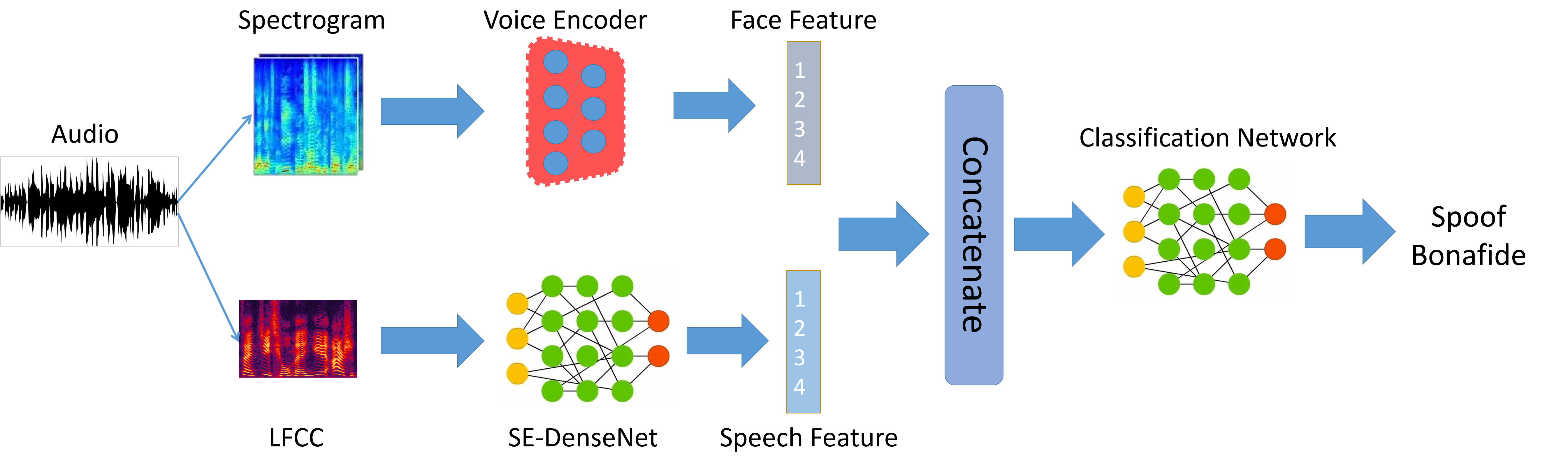}
	\caption{Model structure. This model is for LA. The structure of our model mainly includes two branches. In the first branch, We extract the spectrogram of speech. And then, we use the voice encoder module to turn the spectrogram into face feature. In the second branch, we extract the linear cepstral representation of speech (LFCC). The SE-DenseNet module is used to get the speech feature. The input of SE-DenseNet is LFCC. Last, the features got by the two branches is concatenated and put into a classification network to complete the spoofing attack detection.}\label{fig:demo01}
\end{figure*}
\begin{figure*}[h]
	\centering
	\includegraphics[scale=0.17]{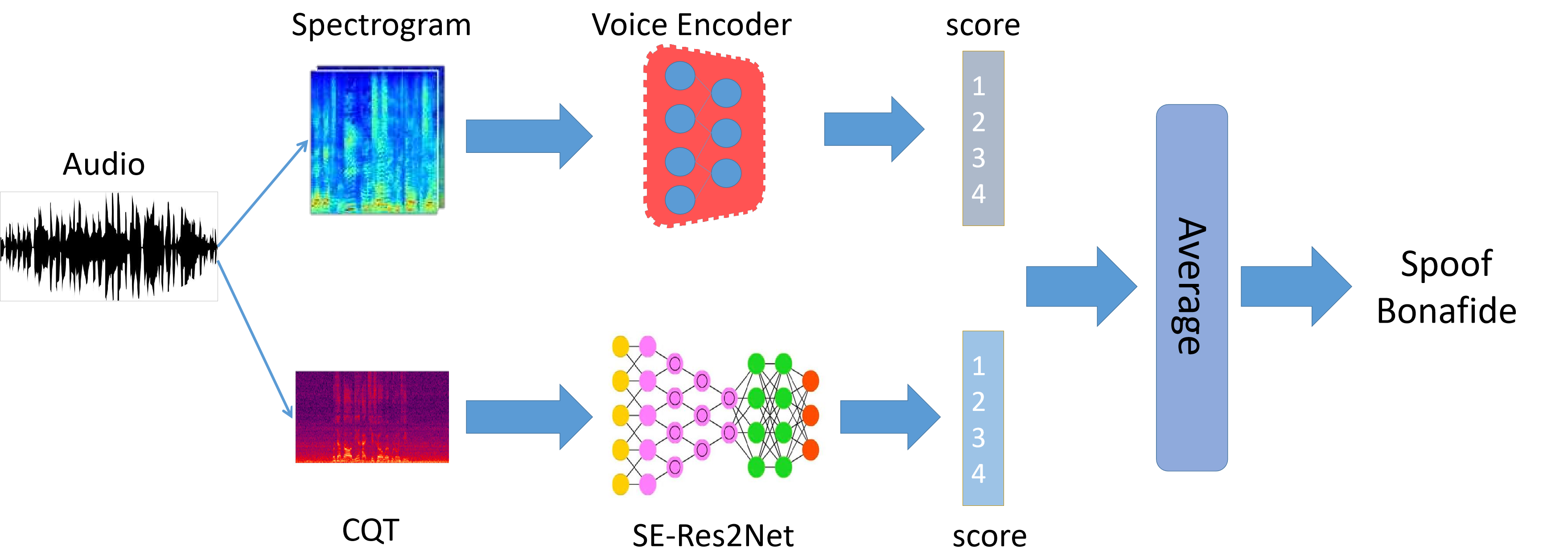}
	\caption{Model structure. This model is for PA. The structure of our model mainly includes two branches. In the first branch, We extract the spectrogram of speech. And then, we still added a voice encoder module, which can use face features to score input speech.  In the second branch, we extract the Constant-Q Transform (CQT). We then send it to the SE-Res2Net module, which scores the input speech using speech features. Finally, we use a weighted average to get the final score. This final score can identify the spoofing attacks.}\label{fig:demo02}
\end{figure*}
We designed two different models for PA and LA respectively. The main idea of the two models is the same, which is to recognize spoofing speech by fusion of physiological and physical features. But the two models are fused in different ways. For LA, front-end fusion is adopted, and its model structure is shown in Figure 1. It consists of four modules, which are the speech processing module, the face feature extraction module, the speech extraction module, and the classification module. For PA, back-end fusion is adopted, and its model structure is shown in Figure 2. It is composed of four modules, namely, speech processing module, face feature classification module, voice feature classification module and fusion module. We will introduce these modules in detail in this section.

\subsection{Problem Definition}

Our goal is to use the information contained in the audio data to predict whether the speech is genuine or spoof. The problem is defined as follows: given an audio file  $ A $, our goal is to predict whether $ A $ is a real speech ($ y=0 $) or a spoofing speech ($ y=1 $), to determine $ A\longrightarrow y(0,1) $.

\subsection{Speech Processing}
In a natural speech, there are great changes in accents, speed, noise and so on, it makes feature extraction of speech very important. Some important information may be lost by using traditional methods, such as MFCC and LPCC. In this paper, we use three different speech processing methods, namely spectrogram, CQT, and LFCC. A large amount of information related to speech presentation features is shown in the spectrogram , so we use it processed the speech to extract face feature. LFCC can capture more spectral details in the high frequency region of the speech spectrum, so we use LFCC processed speech to extract deep audio feature. CQT is a time-frequency representation, where the frequency bins are geometrically spaced, and the q factor (the ratio of the center frequency to the bandwidth) of all bins is equal. This means that the frequency resolution of the low frequency is better, and the time resolution of the high frequency is better.

We use the STFT \cite{durak2003short} to get a spectrogram of a speech sample signal. A large amount of information related to the feature of speech statements is shown in the spectrogram, which combines the feature of spectrum and time waveform. It can show how the speech spectrum changes over time. It also contains a lot of space and time information, which can accurately represent the sound waveform. This will be very helpful for our later modules. In this module, we first frame the audio sample and divide it into small frame sequences. Framing is a necessary task, because we use the STFT when we deal with speech signals, which requires that the audio signal be stationary. The frequency of speech signals is constantly changing. But in a  short period of time we can assume that the frequency of speech signals is fixed. So we can use the audio after framing. Then we use the Hanmming window \cite{podder2014comparative} to window the audio and set the window length to 400. Using the Hanmming window, multiplying each frame signal with a smooth window function can reduce the spectrum energy leakage and reduce the error caused by the spectrum leakage. Finally, the speech signals that have no periodicity show some characteristics of periodic function. We take the STFT of each frame,  the STFT is defined as:
$$ X(n,w)=\sum\nolimits_{m=- \infty}^\infty x(m)w(n-m)e^{-jwm} \eqno{(1)} $$ 
where $ x(m) $ is the input signal, $ w(m) $ is the window function, it's reversed in time, and it has $ n $ sample offsets. $ X(n,w) $ is a two-dimensional function of time $ n $ and frequency $ w $, it connects the time domain and the frequency domain of the signal. Then we transform the signal as:
$$ S(n,w)=\lvert X(n,w) \rvert ^{0.3}  \eqno{(2)}$$
we splicing the Angle values of $ X(n,w) $ and $ S(n,w) $ on dimension two to get the spectrogram.

The LFCC is a form of cepstrum, known as linear cepstrum representation of speech. It is the result of taking the logarithm of the spectrum of the input speech signal and then applying a fourier transform to it, and it is called the linear cepstrum because these coefficients are computed by a linear filter bank which has a better resolution at higher frequencies. To calculate the LFCC, the discrete cosine transform with a log energy filter bank  as shown in (3), where $X_{i}$ is the i-th filter output of the log-energy, $j$ denotes the exponent of the linear frequency cepstral coefficients, $M$ is the number of coefficients to compute and  $B$ is the number of triangular filters.

$$ LFCC_{j}=\sum_{i=1}^B X_{i}cos(j(i-1/2)\pi/B),j=0...M  \eqno{(3)}$$

Constant-Q transform (CQT) refers to a technique that transforms the time domain signal $x(n)$ into the time-frequency domain, so that the center frequencies of the frequency boxes are geometrically spaced, and the Q-factors are equal. In practice, this means that the frequency resolution of the low frequency is better, and the time resolution of the high frequency is better. The CQT transform $X^{CQ}(k, n)$ of a discrete time-domain signal $x(n)$ is defined by
$$ X^{CQ}(k,n)=\sum_{j=n-\lfloor N_{k/2\lfloor}}^{n+\lfloor N_{k}/2\lfloor}x(j)a_{k}^{*}(j-n+N_{k}/2) \eqno{(4)}$$
where $k = 1,2,3\dots ,K$ indexes the frequency bins of the CQT, $\lfloor \cdot \lfloor$ stands for rounding towards negative infinity, and $a_{k}^{*}(n)$ stands for the complex conjugate of $a_{k}(n)$. The basis function $a_{k}(n)$ is a time-frequency atom, also called a complex-valued waveform.

\subsection{Voice Encoder Module}


\begin{table*}[t]\centering \scriptsize \tiny \renewcommand\arraystretch{1.0}  \caption{\label{tab:01}Basic Voice Encoder Architecture} \begin{tabular}{cccccccccccccccc}    \toprule     &  & CONV & CONV & CONV &  & CONV &  & CONV &  & CONV & & CONV & CONV & & AVGPOOL   \\  Layer & Input & BN & BN & BN & MAXPOOL & BN & MAXPOOL & BN & MAXPOOL & BN & MAXPOOL & BN & BN & CONV & BN  \\ &  & RELU & RELU & RELU &  & RELU &  & RELU &  & RELU &  & RELU & RELU &  & RELU   \\ \midrule  Kernel size & - & 4$ \times $4 & 4$ \times $4 & 4$ \times $4 & 2$ \times $1 & 4$ \times $4 & 2$ \times $1 & 4$ \times $4 & 2$ \times $1 & 4$ \times $4 & 2$ \times $1 & 4$ \times $4 & 4$ \times $4 & 4$ \times $4 & $ \infty\times $1   \\  Channels & 2 & 64 & 64 & 128 & - & 128 & - & 128 & - & 256 & - & 512 & 512 & 512 & -  \\   Stride & - & 1 & 1 & 1 & 2$ \times $1 & 1 & 2$ \times $1 & 1 & 2$ \times $1 & 1 & 2$ \times $1 & 1 & 2 & 2 & 1  \\       \bottomrule   \end{tabular}   \end{table*}

Our voice encoder module consists of convolutional neural network and fully connected layer, which converts spectrum of the speech into face feature. This module is independently trained and remains fixed during the experiment. The structure of this module is shown in Table \ref{tab:01}. The blocks of a batch normalization \cite{ioffe2015batch} alternate with max-pooling layers, ReLU and convolution layer. In order to retain the frequency information of speech, it is only pooled along the time dimension of the spectrogram. There are many acoustic features in the frequency information, which can help improve the performance of our model. Finally, we add an average pooling layer. This can not only accommodate different length of speech input, but also better aggregation of information. At the end of the module, we add two fully connected layers to generate 4096-D face features.

We pre-trained this model in a self-supervised manner \cite{hendrycks2019using}. Using the speaker's voice and face in the video as surveillance signals. For this reason, we used the AVSpeech datasets \cite{ephrat2018looking}, which contains a large number of speakers' voices and face images in videos. First, extract a single frame containing the face of the speaker from each video fragment, and input it into the vgg-face model. This model is also a pre-trained model, and it is used to extract the 4096-D feature vector $ v_{f} $. the feature $ v_{s} $ is the supervision signal for our voice encoder, which is trained to predict $ v_{f} $.

To define the loss function, We use the L2 normal to standardize $ v_{f} $ and $ v_{s} $. the L2 normal is defined as:
$$ ||x|| = \sqrt{\sum\nolimits_{i=1}^n x_{i}^{2}} \eqno{(5)}$$
where $ x_{i} $ is a member of the matrix, $ ||x||_{2} $ is the L2 normal of a matrix. We use it for $ v_{f} $ and $ v_{s} $. Then the loss function is defined as:
$$ L=(||v_{f}||_{2}-||v_{s}||_{2})^{2} \eqno{(6)}$$
where $ L $ is the loss function of this model, after that we use it to train our model.

According to the acoustic features of speech in PA and LA, we designed two different voice encoder modules. The network structure of these two modules is roughly the same, but the details are fine tuned. For LA, in order to finally output the 4096-D speech feature, we designed two fully connected layers with the number of channels of 4096 at the end of the network, as shown in Figure 3(a). For PA, our goal is to get the score of the input speech, so we designed two fully connected layers and a softmax function at the end of the network, as shown in Figure 3(b).

\begin{figure}[ht]
	\centering
	\subfigure[Voice Encoder for LA] {
		\label{fig:a}
		\includegraphics[scale=0.14]{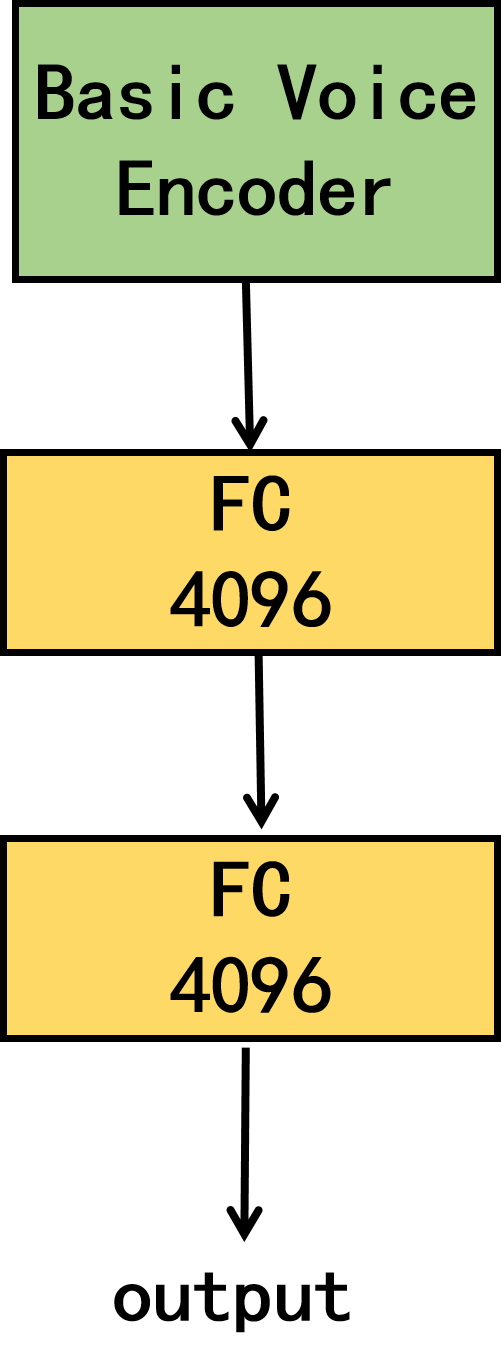}
	}
	\hspace{0.3in}
	\subfigure[Voice Encoder for PA] {
		\label{fig:b}
		\includegraphics[scale=0.14]{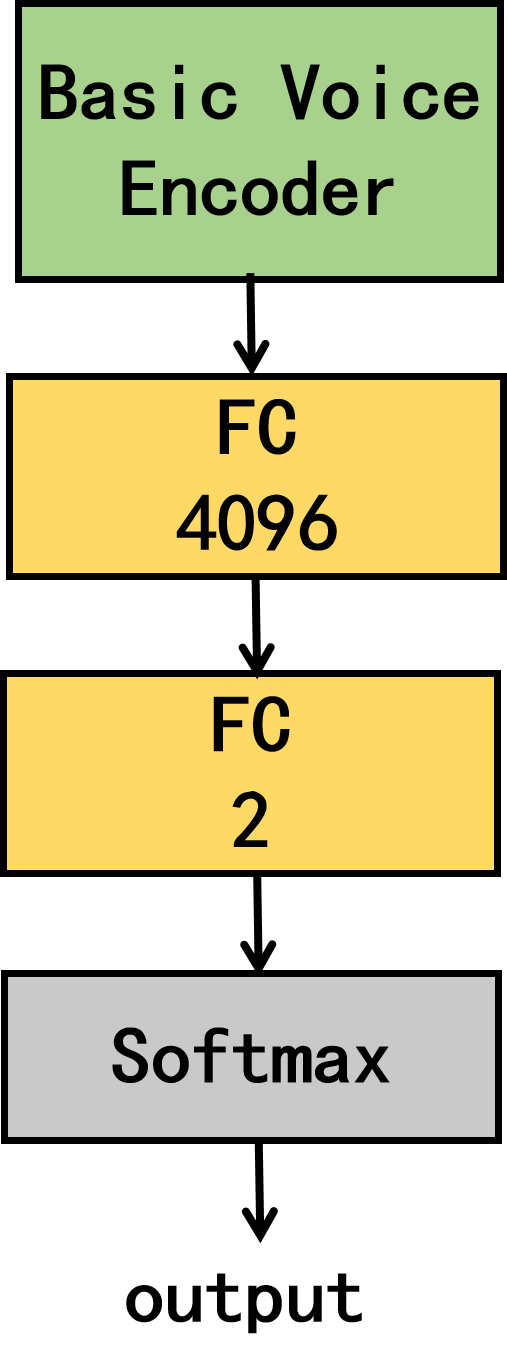}
	}
	
	\caption{ Fine-tuned the Voice Encoder module }
	\label{fig}
\end{figure}

\subsection{SE-DenseNet Module}
\begin{figure*}[h]
	\centering
	\includegraphics[scale=0.13]{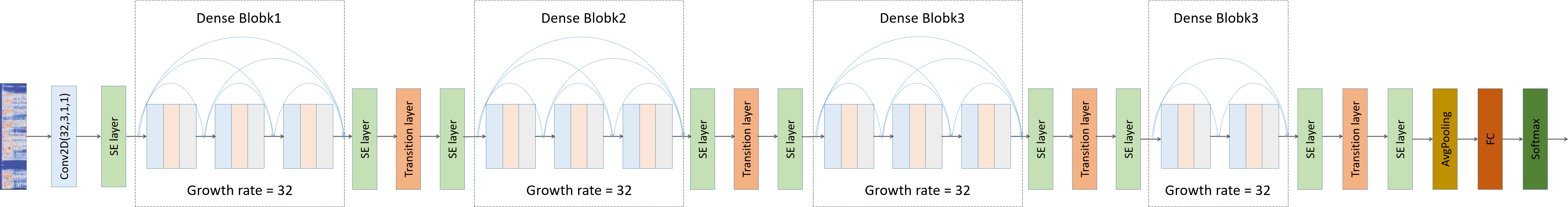}
	\caption{SE-DesneNet network structure}
	\label{fig:pathdemo}
\end{figure*}

The overall structure of our proposed network model is shown in Figure 4. the network consists of 11 regular convolutional layer groups, each of which consists of three parts: convolution, batch normalization , and leaky-ReLU. Moreover, it has a standard convolutional layer, four transition layers, eight squeeze and excitation layers, one average-pooling layer, two fully connected layers, and one softmax layer.

The filter size of the first standard convolution layer is 3×3, with stride and padding one, and outputs 32 feature maps. Four dense block in this structure. Each dense block structure is shown in Figure 5. Densely connections are introduced in each dense block. For the convolutional layers in each block we take the feature maps of all previous layers as input. Three convolutional layer groups are used in the first three dense block and two convolutional layer groups in the last dense block. The receptive field of the four dense blocks is 3 × 3. The growth rate of all four dense block is 32, which means that each convolutional layer outputs 32 feature maps in the block. The transition layer after each block is designed as a convolution of 1 x1, which is designed to reduce the number of input feature maps. SE layer is composed of global average-pooling layer, fully connection layer, ReLU layer and, sigmoid function. By multiplying the output of the global average-pooling layer with the output of the fully connected layer, the SE layer output is obtained. The output of the global average-pooling layer is sent to the full connection layer. Finally, the distribution of two kinds of labels is generated through the softmax layer.

The goal of the ASVspoof challenge is to calculate the countermeasure (cm) score for each input audio file. High cm scores represent bona fide speech, while low cm scores represent spoofing attacks. Calculated from the softmax output using the log-likelihood ratio:
$$ CM(s)=log(p(bonafide|s;\theta))-log(p(spoof|s;\theta))   \eqno{(7)}$$
where $s$ is the audio signal under test and $\theta$ represents the model parameters, $spoof$ represent spoofing attacks, $bonafide$ represent bona fide speech.

\subsubsection{Dense Connectivity}
\begin{figure}[h]
	\centering
	\includegraphics[scale=0.2]{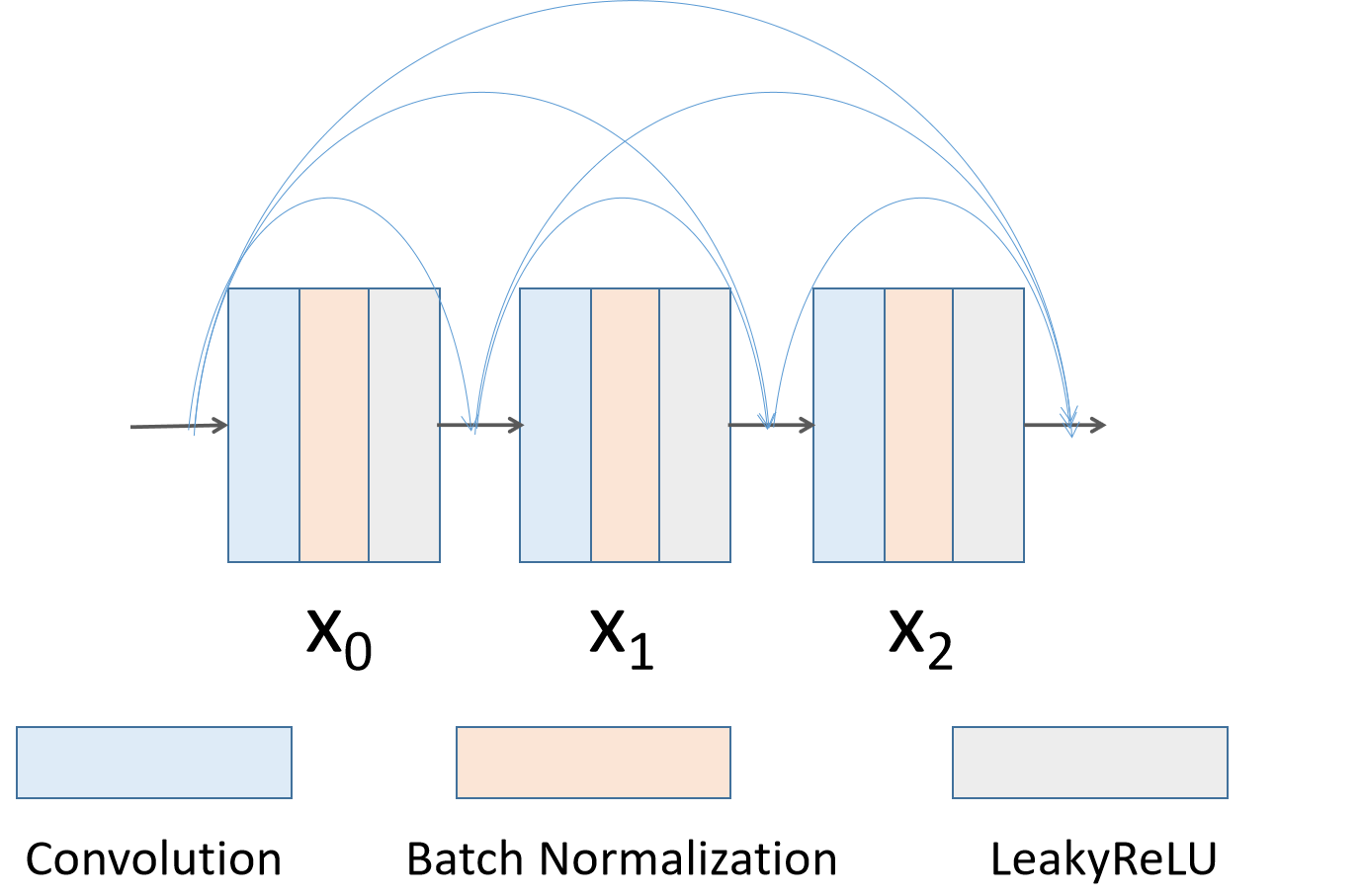}
	\caption{Ilustration of the dense connectivity in dense block.}
	\label{fig:pathdemo}
\end{figure}
Dense connectivity is introduced to our model inspired by \cite{huang2017densely}. In this mode, the output of each convolutional layer is connected to all subsequent convolutional layers, which all have the same size feature map. As shown in Figure 4, the $l^{th}$ layer receives the feature maps from all previous layers, $x_{0},x_{1},...,x_{l-1}$, as the input:
$$ x_{l}=H_{l}([x_{0},x_{1},...,x_{l-1}])   \eqno{(8)}$$
where $x_{0},x_{1},...,x_{l-1}$ refers to the concatenation of the feature maps produced in layer $0,...,l\raisebox{0mm}{-}1$. $H_{l}(\cdot)$ is a composite function of operations including batch normalization, leaky-ReLU and convolution.

Propagate what is learned at the different layers back so that input at the subsequent layers is dynamic and nondeterministic. This ensures the maximum flow of information between network layers. With this densely connected mode, some audio features that are only contained in the previous layer can be transferred to the later layer. This allows the network to learn more useful information. This also enhances the transmission of audio feature. In addition, densely connected mode has many advantages when it comes to back propagation. For example, gradients can bypass activation functions and flow directly from the lower layer to the previous layer, generating implicit deep supervision \cite{huang2017densely}. This can effectively alleviate the problem of gradient disappearance and make the network easier to train.Compared with L-layer traditional convolutional neural networks with L connections, the dense connectivity introduces $\frac{L+1}{2}$ connections. This connection does not require relearning redundant feature mappings. In the convolutional network, the dense model has better parametric efficiency than the traditional model.

\subsubsection{Squeeze and Excitation Blocks}
\begin{figure}[h]
	\centering
	\includegraphics[scale=0.5]{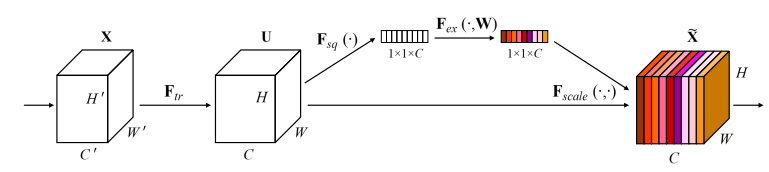}
	\caption{A Squeeze-and-Excitation block}
	\label{fig:pathdemo}
\end{figure}
The squeeze and excitation block can be added to any given structure.It's actually a computational cell. To describe the squeeze and excitation block, we first define the concept of convolution:
$$F_{tr}:X\rightarrow U,X\in R^{H'\times W' \times C'},U\in R^{H\times W \times C}  \eqno{(9)}$$
where $X$ represents the input matrix, $U$ represents the output matrix, $H,W,C$ represents the three dimensions of the matrix. Let $V=[v_{1},v_{2},...,v_{C}]$ represent the learned filter kernel set, where $v_{c}$ represents the parameter of the $c-th$ filter. We can then write the outputs of $F_{tr}$ as $U=[u_{1},u_{2},...,u_{C}]$, at this time $u$ can be expressed as:
$$ u_{c}=v_{c}*X =\sum_{s=1}^{C'} v_{c}^{s}*x^{s} \eqno{(10)}$$
where $v_{c}=[v_{c}^{1},v_{c}^{2},...,v_{c}^{C'}]$, $X=[x^{1},x^{2},...,x^{C'}]$, and $*$ represents convolution, while $v_{c}^{s}$ is a two-dimensional space core, so it represents a single channel of $v_{c}$ acting on the corresponding channel of $X$. Since the output is generated by the sum of all channels passed, the channel dependencies are implicitly embedded in $v_{c}$, but these dependencies are entangled with the spatial correlation captured by the filter. Our goal is to enable the network to hold more useful feature information. This allows subsequent steps to take advantage of this information and discard some less useful feature information. We do this through interdependencies between channels, recalibrating the filter response by squeeze and excitation them in two steps before they are fed into the next transformation. An SE building block construction is shown in Figure 6.

\subsection{SE-Res2Net Module}

\begin{figure}[h]
	\centering
	\includegraphics[scale=0.12]{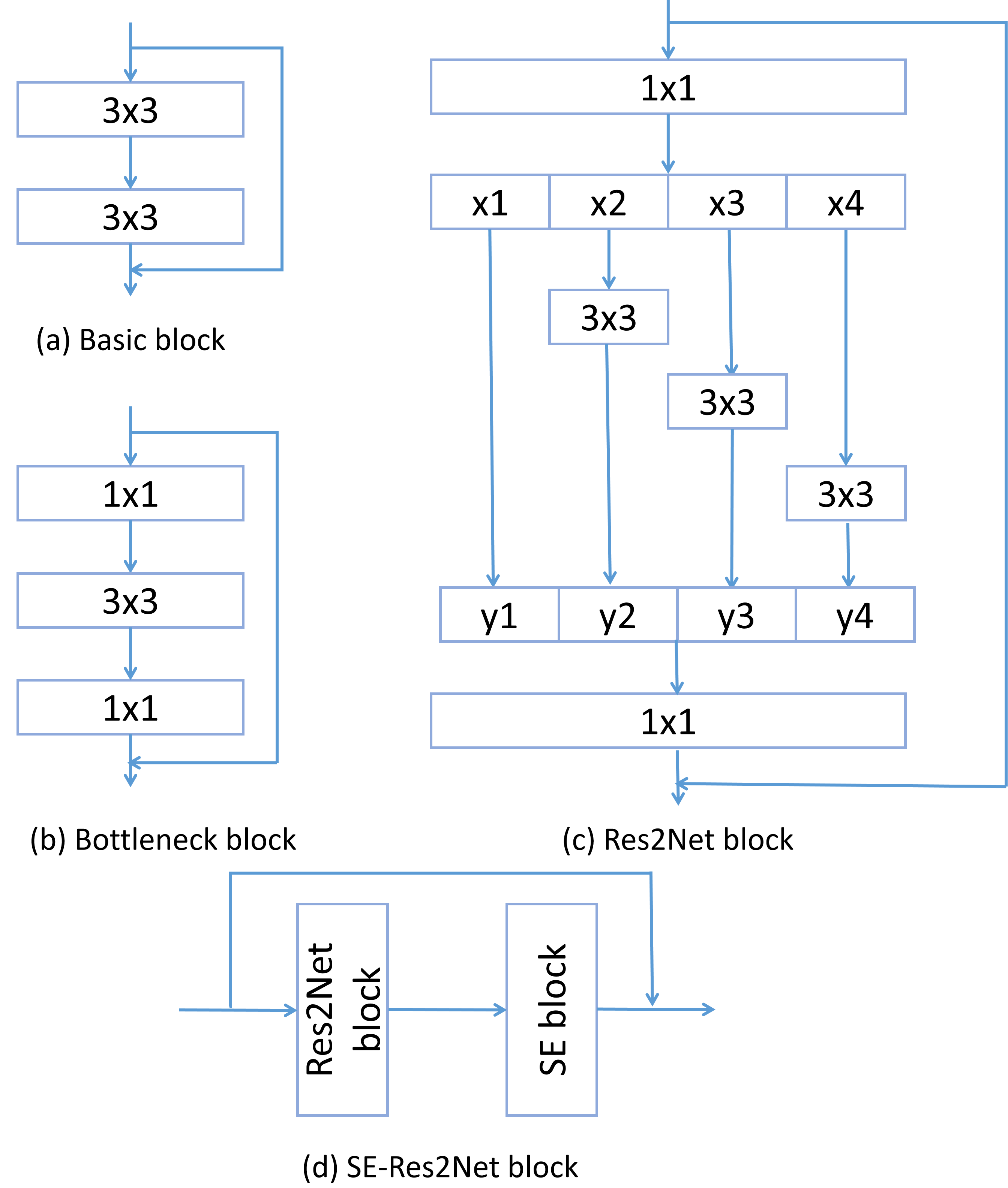}
	\caption{Structure of basic block, bottleneck block, Res2Net block(scale dimension $s = 4$), and SE-Res2Net block}
	\label{fig:pathdemo}
\end{figure}

SE-Res2Net module is implemented by SE Block through Res2Net network integration. This network architecture aims to improve multiscale representation by increasing the number of acceptance domains available. A comparison of base blocks, bottleneck blocks, and Res2Net blocks is shown in Figure 7. Res2Net blocks are obtained by deforming bottleneck block, while SE-Res2Net block are obtained by connecting Res2Net block to SE block.

The structure of Res2Net block is shown in Figure 7(c). After a $1\times1$ convolution. The input feature map is evenly divided into $s$ subsets according to the channel dimension, expressed in $x_{i}$, where $i\in\{1,2,\dots,s\}$. Every $x_{i}$ is processed by a $3\times3$ convolution filter $K_{i}()$, except $x_{1}$.  Starting from $i=3$, $x_{i}$ is added with the output of $K_{i-1}$ before being fed into $K_{i}()$. This process can be expressed as:
$$ y_{i}= \begin{cases}
	\ x_{i} & i=1\\
	\ K_{i}(x_{i})& i=2 \\
	\ K_{i}(x_{i}+y_{i-1})& 2<i\le s \\
\end{cases} \eqno{(11)} $$

where $s$ represents the scale dimension, which represents the number of partitions applied to the segmentation feature map. This hierarchical residual connection can realize multi-size receptive fields within a block, thereby generating multiple feature scales. Finally, it connects all the divided blocks and passes them through an $1\times1$ convolution filter to maintain the channel size of this residual block.

In Res2Net block, segmentation is processed in a multi-scale manner, which is conducive to extracting global and local information. In order to better fuse the fusion information at different scales, we spliced all the segments together and passed $1\times1$ convolution. Segmentation and connection strategies can enforce convolution to handle features more efficiently. In order to reduce the number of parameters, we omitted the first split convolution, which can also be seen as a form of feature reuse. We implement hierarchical residual connections between filter groups, which also can effectively reduce the number of model parameters. Here, we use $s$ as the control parameter of the scale dimension. When the threshold of $s$ becomes larger, we can learn more abundant features of the size of the receiving field. Since the connection operation requires almost no calculation, the computational overhead introduced by the connection can be ignored.

SE block recalibrates the features response of the channel adaptively by explicitly modeling the interdependence between the channels. We connect an SE block after the Res2Net block to form an SE-Res2Net block, as shown in Figure 7(d). Our experimental results show that this connection method can further improve spoofing attack detection.

\subsection{Classification Module}

\begin{table}[t]\centering\scriptsize    \caption{Classification Module Architecture}\label{classification}
	
		\begin{tabular}{ccc}    \toprule    Layers & Architecture  & Fusion Feature  \\ \midrule  Convolution & 3 $\times$3,stride 1 & 64$ \times $66  \\  SE Block & - & 64$ \times $66  \\  Dense Block1 & (3 $\times$ 3) $\times$ 3 & 64$ \times $66 \\  SE Block & - & 64$ \times $66  \\  Transition layer & 1 $\times$ 1 conv & 64$ \times $66 \\    Max pooling & 2 $\times$ 2 max pool, stride 2 & 32$ \times $33 \\  Dense Block2 & (3 $\times$ 3) $\times$ 3 & 32$ \times $33 \\   SE Block & - & 32$ \times $33  \\ Transition layer & 1 $\times$ 1 conv & 32$ \times $33 \\ Average pooling & 2 $\times$ 2  average pool, stride 2 & 16$ \times $ 16 \\ Dense Block3 & (3 $\times$ 3) $\times$ 3 & 16$ \times $ 16 \\  SE Block & - & 16$ \times $16  \\ Transition layer & 1 $\times$ 1 conv & 16$ \times $ 16 \\   Max pooling & 2 $\times$ 2 max pool, stride 2 & 8$ \times $8 \\ Dense Block4 & (3 $\times$ 3) $\times$ 2 & 8$ \times $ 8 \\   SE Block & - & 8$ \times $8  \\  Transition layer & 1 $\times$ 1 conv & 8$ \times $8 \\ \multirow{2}{*}{Classification layer} & global average pool & 1$\times$1 \\  & Dropout, 128 FC, softmax & - \\ \bottomrule   \end{tabular}  \end{table}

The classification module is mainly prepared for logical scenarios, because logical scenarios need a classification module to distinguish spoofing attacks after feature fusion. Our classification network is a variant of DenseNet. It mainly consists of dense block, squeeze and excitation block, transition layer and fully connected layer.  This module is trained after the other modules.

The input to our classification module is the fusion feature. It's a combination of physical features and physical features. Through the Voice Encode module, we can get the 4096-D face features , which contains some information related to the speaker's face, and we call it physiological features. With the SE-Densenet module, we can get the speech feature of 128-D, which contains the information about the speaker's voice, and we call the physical features. Then we use the splicing function to splice these two features to form our fusion feature.

The overall structure of the classification module is shown in Table \ref{classification}.  It consists of a convolution layer, four dense blocks, four transition layers,squeeze and excitation block, two max pooling layers, an average polling layer and two fully connected layers. The first convolutional layer has a convolutional kernel size of 3x3 and a step size of 1. The first three dense blocks contain three convolutional layers with 3x3 convolutional cores; the last dense block only contains two convolutional layers, and the convolutional cores are also 3x3. All transition layers in this module are convolutional layers of 1x1, which are placed behind each dense block. In order to better transmit network information backwards, we set the growth rate of each dense block as 32, which means that each convolutional layer will output 32 feature maps. We add a pooling layer behind each transition layer to prevent unnecessary parameters from adding time complexity and to increase feature integration. After the first transition layer and the third transition layer, we add the max pooling layer, which can ensure the invariance of translation and rotation of the feature map, and the receptive field of the current feature map can also be increased. We add the average pooling layer after the second transition layer, which can reduce the size of spatial information, reduce the number of network parameters and reduce the risk of overfit. After the last transition layer, we add a global average pool layer, which can pool the feature map of the last layer by an average value of the whole image to form a feature point, and these feature points can be used to form the final feature vector. The last two fully connected layers are used for classification.

\subsection{Concatenation and  average strategy}
In the logical scenario, we use the concatenation strategy. We use this strategy to fuse face features with speech features. In the previous module, we have successfully extracted 128-D speech features and 4096-D face features. In this module, we connect the voice features to the back of the face features to form the 4224-D fusion features. In order to enhance the receptive field of fusion features, we add a dimension to the fusion features, turning one-dimensional fusion features into two-dimensional fusion features. Once this is done, we have completed the concatenation strategy for the logical scenario.

In the physical scenario, we used the averaging strategy. This strategy can be used to synthesize the scores obtained in previous modules into the final score. In this method, we evaluated various averaging strategies and finally found that the weighted averaging strategy was most suitable for physical scenarios, so we finally chose the weighted averaging method. Because of the large dimension of face features, we give it a smaller weight. The dimension of the speech feature is small, so we give it bigger weight. Finally, after adjustment, we determine the weight of face features is 0.1, and the weight of speech features is 0.9.

\section{Experiments}

\subsection{Datasets}

\begin{table}[h] \centering \scriptsize    \caption{ The number of speakers and the number of speechs in the development set and training set, in the ASVspoof 2019 database.}\label{tab:03}
	
	\begin{tabular}{ccccccc}    \toprule     & \multicolumn{2}{c}{Speakers}  & \multicolumn{4}{c}{Utterances}  \\ 
		\multirow{2}{*}{Subset} & \multirow{2}{*}{Female} & \multirow{2}{*}{Male} & \multicolumn{2}{c}{Physical access} &  \multicolumn{2}{c}{Logical access} \\
		& & & Spoof &  Bona fide  & Spoof & Bona fide \\ \midrule
		Training & 12 & 8 &  22800 & 2580  & 48600 & 5400  \\
		Development & 12 & 8 & 22296 & 2548  & 24300 & 5400  \\ \bottomrule   \end{tabular}  \end{table}

The ASVspoof 2019 data set contains two partitions for the evaluation of LA and PA scenarios. They are both from the VCTK basic corpus, which includes voice data captured from 107 speakers (46 males, 61 females). The LA and PA databases themselves are divided into three data sets, namely training, development and evaluation, which include 20 (8 males, 12 females), 10 (4 males, 6 females) and 48 ( (21 males, 27 females) speakers' speeches. The data of these three sets are not duplicated. In both the development set and the evaluation set, spoofing attacks from the training set using the same algorithm occur (designated as known attacks). Spoofing attacks that have never been seen before will also appear in the evaluation set, and they are generated using different algorithms (designated as unknown attacks). The number of bona fide speech and spoofing speech in the development set and training set is shown in Table \ref{tab:03}. Therefore, reliable spoofing detection performance requires that the system can well summarize previously invisible spoofing attacks.

ASVspoof 2019 is a datasets for anti-spoofing research for automatic speaker verification. ASVspoof 2019 has been the third version of the datasets, which mainly contains three kinds of spoofing attacks: TTS, VC and replay attack. In LA, TTS, and VC are the main way of attack, however, PA primarily contains physical attack. We will design two different strategies for resisting spoofing attacks against these two different attack methods. This datasets divides the data into three parts: training set, development set and evaluation set. In the training set, 17 different TTS systems and the VC systems were used to generate spoofing speech, and the speech generated by these different systems made the experiment more robust. Spoofing speech generated by 17 different systems was also used in the development set, but only 6 of these systems have been used in the training set, and the other 11 systems have not been used. In the evaluation set, they even use eleven unknown systems and two known systems. These systems include many advanced systems,  such as generation of countermeasure network, neural waveform model, waveform splicing, waveform filtering, etc . This makes the data sample of this datasets extremely rich, and the experimental results have high credibility.

\subsubsection{Logical access}


The LA data set contains spoofing voice and bona fide voice data generated using 17 different VC and TTS systems. The data used by the TTS and VC systems to generate false speech comes from the VCTK database. The speech data in the logical data set does not overlap with the data in the previous version of the ASVspoof data set. Among the 17 systems in LA, 6 systems were designated as known attacks, and the other 11 systems were designated as unknown attacks. Only known attacks are included in the training set and development set, and both known attacks and unknown attacks are included in the evaluation set. Among the 6 known attacks, there are 2 VC systems and 4 TTS systems. The TTS system uses neural network-based speech synthesis or waveform cascade, using WaveNet-based vocoder \cite{oord2016wavenet} or traditional source filtering vocoder \cite{2016WORLD}. The VC system adopts methods based on neural network and spectral filtering \cite{2006Effect}. These 11 unknown systems are composed of 2 VCs, 6 TTSs and 3 hybrid TTS-VC systems, which are implemented using various waveform generation methods, including GriffinLim \cite{griffin1984signal}, classical speech coding, , neural waveforms model \cite{oord2016wavenet, wang2019neural}, generative confrontation network \cite{tanaka2018synthetic}, waveform filtering \cite{kobayashi2014statistical}, waveform cascading, spectral filtering and their combination.

\subsubsection{Physical access}

{The bona fide speech and spoofing speech contained in PA are generated by recording the microphone of ASV system in real environment. It is also divided into three subsets: training set, development set and evaluation set.Training and development data are created based on 9 different playback configurations and 27 different acoustics. The acoustic configuration includes a detailed combination of 3 types of speaker to asv microphone distance, 3 types of reverberation, and 3 types of room sizes. The playback configuration includes three types of attacker-to-speaker recording distances and three types of speaker quality. A random attacker-to-talker recording distance and random speaker quality corresponding to a given configuration category are used to simulate a replay attack. The room size, the reverberation level and the distance from the speaker to the asv microphone for bona fide and spoofing speech are all random.}

{The data generation method of the evaluation set is the same as the generation method of the development set and the training set, but the replay configuration and random acoustics of the evaluation set are different. The room size, the distance between the speaker and the asv microphone, the reverberation level, the recording distance between the attacker and the speaker, and the speaker quality settings are all different from the training and development set settings. Although the configuration categories of these sets are the same and the setting parameters are known, the spoofing voice used by the specific playback device to simulate is different and unknown. So even if the system achieves good results in the training and development set, it may not perform well in the evaluation set. This requires our system to have good robustness.}

\subsection{Baseline Model}

\begin{table}[h]\centering \scriptsize   \caption{ \label{tab:04} t-DCF and EER results for two baseline countermeasures and both logical and physical access scenarios.}
	
	\begin{tabular}{ccccc}    \toprule     & \multicolumn{2}{c}{Logical access}  & \multicolumn{2}{c}{Physical access}  \\ 
		& t-DCF & EER(\%) &  t-DCF & EER(\%) \\ \midrule
		LFCC-GMM & 0.0663 & 2.71 & 0.2554 & 11.96  \\
		CQCC-GMM & 0.0123 & 0.43 & 0.1953 & 9.87  \\ \bottomrule   \end{tabular}  \end{table}

The organizers provided two baseline models, They are based on two different acoustic features, namely linear frequency cesptral coefficients (LFCC) and constant Q cepstral coefficients (CQCC), and they are both use a GMM binary classifier. The results of these two approaches are shown in Table \ref{tab:04}. 

\subsubsection{LFCC-GMM}
First, the power spectrum is integrated by overlapping band-pass filter, the power spectrum is logarithmic compressed and the cepstrum coefficients are obtained by discrete cosine transform (DCT). A rectangular window was used for integration, and a filter extracted LFCC according to a linear scale interval, then a gaussian mixture model was used as a classification network.

\subsubsection{CQCC-GMM}
Instead of the traditional fourier transform, the constant Q transform (CQT) is used, where the frequency multiplier of CQT is distributed geometrically, and the center frequency of each filter is linearly spaced. Its frequency spectrum is nonlinear, the center frequency is distributed exponentially, the filter bandwidth is different, but the center frequency and bandwidth ratio is constant Q filter bank. This avoids the disadvantage of uniform time-frequency resolution. Finally,  the gaussian mixture model is used as a classification network.

\subsection{Evaluation Metrics}

{ASVspoof 2019 has two evaluation metrics. Among them, tandem detection cost function (t-DCF) is the main evaluation metrics, equal error rate (EER) was a secondary evaluation metrics. Two indicators, t-DCF and EER, were calculated using test scores. Each number corresponds to an audio file. We should assign a real value to each trial, a finite number that reflects the support of the two conflicting assumptions that the trial was real or spoofed by the audio. Similar to the previous two versions of ASVspoof, a high detection score should indicate goodwill and a low detection score should indicate spoofing attacks.}

\subsubsection{t-DCF}

{t-DCF is the main evaluation indicator of ASVspoof 2019. It is an evaluation method of a series system, which requires the use of both the spoofing countermeasure (CM) (designed by the participant) and the ASV system (provided by the organizer). The performance of the two combined systems is evaluated by a formal minimum normalized series detection cost function (t-DCF). t-DCF is defined as follows:}
{$$ t\raisebox{0mm}{-}DCF_{norm}^{min}=\mathop{min}\limits_{s}\{ \beta P_{miss}^{cm}(s)+P_{fa}^{cm}(s) \}  \eqno{(12)}$$ }
{where $\beta$ depends on the performance of the ASV system and the parameters of the anti-spoofing strategy. , while $ P_{miss}^{cm}(s) $ and $ P_{fa}^{cm}(s) $ are the CM miss alarm rate and false alarm rate under the threshold s. The minimum value in (12) is to use a known key value on all thresholds of the given data (development or evaluation), which corresponds to the oracle calibration. When other conditions remain the same, $ \beta $ is inversely proportional to the ASV false acceptance rate of a specific attack: when the CM incorrectly rejects bona fide speech, the penalty is higher when the attack efficiency is low. Similarly, for more effective attacks, when the CM mistakenly accepts spoofs, the relative penalty is higher.
}
\begin{figure}[h]
	\centering
	\includegraphics[scale=0.5]{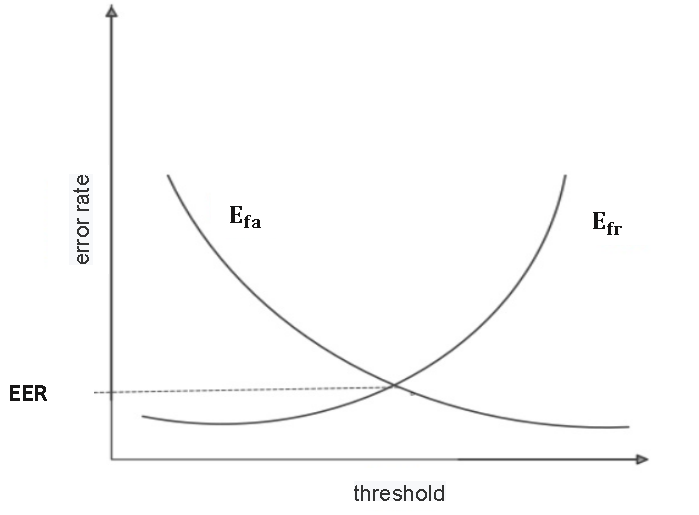}
	\caption{Error rate changes with the threshold.}
	\label{fig:pathdemo}
\end{figure}
\subsubsection{EER}
EER is a secondary evaluation metric, which was used as the primary evaluation metric in the previous two versions of ASVspoof. EER corresponds to the CM threshold. At this threshold, the false alarm rate and the miss alarm rate are equal. Same as t-DCF, the calculation of EER is also generated from the scoring file of our CM system. The CM operating point corresponding to EER has the same false alarm rate and miss alarm rate, which is the main measurement of the previous version of ASVspoof. If there is no clear link to the impact of CMs on the reliability of the ASV system, EER may be more suitable as a measure of spoofing audio detection, that is, in the absence of an ASV system.

EER is associated with false rejection rate $E_{fr}$ and false acceptance rate $E_{fa}$. We first define false rejection rate $E_{fr}$ and false acceptance rate $E_{fa}$:
$$ E_{fr} = \frac{N_{fr}}{N_{target}}  \eqno{(13)}$$
$$ E_{fa} = \frac{N_{fa}}{N_{non-target}}  \eqno{(14)} $$
where $N_{fr}$ and $N_{fa}$ refer to the number of false rejections and false acceptances in the test, respectively. $N_{target}$ and $N_{non-target}$ respectively refer to the total number of bona fide speech and spoof speech in the test. When the threshold in the system is fixed, $E_{fr}$ and $E_{fa}$ are fixed. As the threshold decreases, more tests will be accepted, with $E_{fa}$ increase and $E_{fr}$ decrease; On the contrary, when the threshold value increase, the test will not be easy to pass, $E_{fa}$ decrease, $E_{fr}$ increase. The change of error rate with threshold value is shown in Fig 8. EER is the error rate when $E_{fr}$ and $E_{fa}$ are equal, and is defined as: $EER=E_{fr}=E_{fa}$.

\subsection{Experimental setup}


For the logical scenario, we trained the voice encoder module of the first branch using ADAM as the optimizer and set the learning rate to 0.001. In the last layer of voice encoder, we use a full connection layer to output 4096-D face features. When training the SE-DenseNet module of the second branch, we still use ADAM as the optimizer and set the learning rate to 0.0005. After 200 epochs of training, 128-D face features were obtained. In the final classification module, we use binary cross entropy as the loss function and log-softmax as the activation function in the last layer to score the input features.

For physical scenarios, we trained the voice encoder module differently. Since the voice encoder module directly outputs the score obtained by face features in the physical scenarios, we directly add two full connection layers and softmax functions after the voice encoder module to score the speech. The optimizer and learning rate in the voice encoder module are the same as in the logical scenario. When training SE-Res2Net, we also used ADAM as the optimizer, and the learning rate was set to 0.0003. After 20 epochs of training, scores based on the assessment of speech features were obtained.

\subsection{Experimental Results}

\begin{table}[h]
	
	\scriptsize
	
	\centering

	\caption{Performance comparison of SE-DenseNet to DenseNet on the ASVspoof 2019 LA evaluation set.}\label{tab05}

		\begin{tabular}{ccc}
			
			\toprule
			
			\multirow{2}{*}{Model} & \multicolumn{2}{c}{Logical Access}  \\
			
			\cmidrule(r){2-3}
			
			&    EER(\%)   &  t-DCF   
			
			\\
			
			\midrule
			
			LFCC+DenseNet  &  4.74  &   0.1245              \\
			
			LFCC+SE-DenseNet  &  3.49  & 	0.0914     			\\
			
			\textbf{LFCC\&Face+SE-DenseNet} & \textbf{2.82} & \textbf{0.0742} \\
			\bottomrule
			
	\end{tabular}

\end{table}

\begin{table}[h]
	
	\scriptsize
	
	\centering

	\caption{Performance comparison of SE-Res2Net to Res2Net on the ASVspoof 2019 PA evaluation set.}\label{tab06}
	
		\begin{tabular}{ccc}
			
			\toprule
			
			\multirow{2}{*}{Model} & \multicolumn{2}{c}{Physical Access}  \\
			
			\cmidrule(r){2-3}
			
			&    EER(\%)   &  t-DCF   
			
			\\
			
			\midrule
			
			CQT+Res2Net  &  1.26  &   0.0312              \\
			
			CQT+SE-Res2Net  &  0.93  & 	0.0254     			\\
			
			\textbf{CQT\&Face+SE-Res2Net} & \textbf{0.85} & \textbf{0.0230} \\
			\bottomrule
			
	\end{tabular}

\end{table}

We will evaluate the effectiveness of our model to audio spoofing detection in this section. In this section, we will further divide into two sections to discuss the performance of our model on LA and PA respectively.

\subsubsection{Logical Access Results}
\begin{table}[h]
	
	\scriptsize
	
	\centering

	\caption{Performance comparison of some known state-of-the-art single models to our proposed model on the ASVspoof 2019 LA evaluation set.}\label{tab07}
		
		\begin{tabular}{ccccc}
			
			\toprule
			
			\multirow{2}{*}{Model} & \multicolumn{2}{c}{Logical Access}  \\
			
			\cmidrule(r){2-3} 
			
			&    EER(\%)   &  t-DCF  
			\\
			
			\midrule
			
			Spec+ResNet+CE\cite{alzantot2019deep}  &  9.68  &   0.2741             \\
			
			MFCC+ResNet+CE\cite{alzantot2019deep}  &  9.33  & 	 0.2042   			\\
			
			CQCC+ResNet+CE\cite{alzantot2019deep}  &  7.69  &   0.2166      		\\
			
			Spec+ResNet+CE\cite{lai2019assert}  &  11.75  &   0.216     		\\

			LFCC+LCNN+A-softmax\cite{lavrentyeva2019stc}  &  5.06  &   0.1000      		\\
			
			FFT+LCNN+A-softmax\cite{lavrentyeva2019stc}  &  4.53  &   0.1028     		\\

			FG-CQT+LCNN+CE\cite{wu2020light}  &  4.07  &   0.102    		\\
			
			Spec+LCGRNN+GKDE-Softmax\cite{gomez2020kernel}  &  3.77  &   0.0842    		\\
			
			Spec+LCGRNN+GKDE-Triplet\cite{gomez2020kernel}  &  3.03  &   0.0776     		\\

			\textbf{Ours: LFCC\&Face+SE-DenseNet+log-softmax} & \textbf{2.82} & \textbf{0.0742}  \\

			\bottomrule
			
	\end{tabular}

\end{table}
We evaluate the effectiveness of audio spoofing detection based on SE-DenseNet model in this section. Results based on DenseNet and SE-DenseNet were used for comparison, as shown in Table \ref{tab05}. Their feature input is speech feature and the fusion feature of speech feature and face feature respectively. Our comparison shows that the SE-DenseNet is superior to the DenseNet in all conditions. This shows that squeeze and excitation Block can improve the spoofing detection performance of the system. t-DCF and EER were both reduced by 26\% using the SE-DenseNet model compared to the DenseNet model. Comparing the SE-DenseNet model using fusion feature with the SE-DenseNet model using only speech feature, we find that the system using fusion feature has significantly better performance than the system using speech feature. This indicates that the face feature extracted is helpful for the system performance. Compared to the SE-DenseNet, which used the speech feature, the EER of the SE-DenseNet using the fusion feature decreased by 19\%. A similar gain can be observed under the t-DCF metric. This further verifies the efficiency of face feature and SE-DenseNet architecture in spoofing attack detection.

We also compare some of the best single systems. SE-DenseNet with the face and speech feature, with some of the best single systems on the ASVspoof2019 LA evaluation sets. Some individual systems that perform well are shown in Table \ref{tab07}, as far as we know. The names of these systems consist of input feature, system structure, and loss functions.

We observed that only a few systems had EER below 4.0\% in the LA evaluation set, and very few systems had good performance on both EER and t-DCF, indicating the challenge of detecting unknown system attacks. Most well-performing systems are exploring powerful model architectures and input features. For LA attacks, the system in \cite{gomez2020kernel}  surpasses all other systems and it is the best single system model in the ASVspoof 2019 Challenge. However, compared to it, the EER of our system was reduced by 6\%, describing the effectiveness of our proposed system.

\subsubsection{Physical Access Results}
\begin{table}[h]
	
	\scriptsize
	
	\centering

	\caption{Performance comparison of some known state-of-the-art single models to our proposed model on the ASVspoof 2019 PA evaluation set.}\label{tab08}
		
		\begin{tabular}{ccc}
			
			\toprule
			
			\multirow{2}{*}{Model}  & \multicolumn{2}{c}{Physical Access} \\
			
			\cmidrule(r){2-3} 
			
			&    EER(\%)   &  t-DCF     
			\\
			
			\midrule
			
			Spec+ResNet+CE\cite{alzantot2019deep}    & 3.81 & 0.0994            \\

			CQCC+ResNet+CE\cite{alzantot2019deep}    & 4.43 & 0.1070    		\\
			
			Spec+ResNet+CE\cite{lai2019assert}     & 1.29 & 0.036   		\\
			
			Joint-gram+ResNet+CE  \cite{Weicheng}                 &  1.23 & 0.0305   		\\
			
			GD-gram+ResNet+CE  \cite{Weicheng}                    &  1.08 & 0.0282   		\\
			
			LFCC+LCNN+A-softmax\cite{lavrentyeva2019stc}    & 4.60 & 0.1053    		\\

			CQT+LCNN+A-softmax \cite{lavrentyeva2019stc}          &  1.23   &  0.0295    		\\

			Spec+LCGRNN+GKDE-Softmax\cite{gomez2020kernel}    & 1.06 & 0.0222   		\\
			
			Spec+LCGRNN+GKDE-Triplet\cite{gomez2020kernel}    & 0.92 &  0.0198   		\\
			
			MGD+ResNeWt+CE   \cite{2019Replay}                              & 2.15  &  0.0465  		\\
			
			CQTMGD+ResNeWt+CE   \cite{2019Replay}                             & 0.94  &  0.0250  		\\

			\textbf{Ours: CQT\&Face+SE-Res2Net+log-softmax}  & \textbf{0.8526} & \textbf{0.0230} \\
			\bottomrule
			
	\end{tabular}

\end{table}
In the physical scenario, we evaluated the performance based on the SE-Res2Net model. The ablation experiment was performed first, and the experimental results are shown in Table \ref{tab06}. We first compare the performance of models that use only physical features CQT+SE-Res2Net with those that use physical and physiological features CQT\&Face+Res2Net. In the experiment, EER of CQT+SE-Res2Net was 0.93\% and t-DCF was 0.0254. While EER and t-DCF of CQT\&Face+Res2Net was 0.85\% and 0.023, respectively. Compared with CQT+SE-Res2Net, EER and t-DCF of CQT\&Face+SE-Res2Net increased by 9\% and 10\%, respectively. This indicates that the physiological features proposed by us are effective for spoofing attack detection. We also conducted a group of experiments with the model without SE block, called CQT+Res2Net. The EER and t-DCF of CQT+Res2Net were 1.26\% and 0.0312, respectively. Compared to the model with SE block added, the performance of CQT+Res2Net is worse. Compared with CQT+Res2Net, CQT+SE-Res2Net had a 35\% increase in EER and 22\% increase in t-DCF. This shows that SE block can still significantly improve the performance of the model in physical scenarios.

We also compared several systems that performed well in physical scenarios, as shown in Table \ref{tab08}. It can be seen from the table that the performance of our model is obviously better than that of other models. Compared to the performance of the best models in the table, our model improved by 10\% and 8\% on EER and t-DCF, respectively.

\section{Conclusion}

According to the characteristics of logical scenario and physical scenario, this paper designs methods to resist spoofing attack for them respectively. For logical scenarios, our method consists of feature extraction, a densely connected network, squeeze and excitation block and feature fusion strategy. For physical scenarios, our method consists of feature extraction, multi-scale residual network, SE block and weighted average strategy. Both methods creatively put forward the strategy of combining physical characteristics with physiological characteristics to resist spoofing attacks. Our experiments show that this strategy is effective for both logical and physical scenarios. We also compared our proposed method with the most advanced DNN based method, and our method achieved better performance in t-DCF and EER metrics. In the logical scenario, our model improved t-DCF and EER by 28\% and 11\%, respectively. In the physical scenario, our model improved t-DCF and EER by 8\% and 10\%, respectively.


%

\appendices



\ifCLASSOPTIONcaptionsoff
  \newpage
\fi



%

\bibliographystyle{IEEEtran}
\bibliography{IEEEexample}

%






\end{document}